\documentclass[runningheads]{llncs}
\usepackage{graphicx}
\usepackage{multicol}
\usepackage{multirow}
\usepackage{amsmath}
\begin{document}
\title{Modeling Helping Behavior in Emergency Evacuations using Volunteer's Dilemma Game}
\titlerunning{Modeling Helping Behavior in Emergency Evacuations}
%
%
\author{Jaeyoung Kwak\inst{1} \and
Michael H. Lees\inst{2} \and
Wentong Cai\inst{1} \and
Marcus E. H. Ong\inst{3, 4}
}
\authorrunning{J. Kwak \textit{et al.}}
%
\institute{
Nanyang Technological University, Singapore 639798, Singapore\\
\email{\{jaeyoung.kwak, aswtcai\}@ntu.edu.sg}\\
\and
University of Amsterdam, Amsterdam 1098XH, The Netherlands\\
\email{m.h.lees@uva.nl}
\and
Singapore General Hospital, Singapore 169608, Singapore\\
\and
Duke-NUS Medical School, Singapore 169857, Singapore\\
\email{marcus.ong.e.h@singhealth.com.sg}
}
\maketitle              
\begin{abstract}
People often help others who are in trouble, especially in emergency evacuation situations. For instance, during the 2005 London bombings, it was reported that evacuees helped injured persons to escape the place of danger. In terms of game theory, it can be understood that such helping behavior provides a collective good while it is a costly behavior because the volunteers spend extra time to assist the injured persons in case of emergency evacuations. In order to study the collective effects of helping behavior in emergency evacuations, we have performed numerical simulations of helping behavior among evacuees in a room evacuation scenario. Our simulation model is based on the volunteer's dilemma game reflecting volunteering cost. The game theoretic model is coupled with a social force model to understand the relationship between the spatial and social dynamics of evacuation scenarios. By systematically changing the cost parameter of helping behavior, we observed different patterns of collective helping behaviors and these collective patterns are summarized with a phase diagram.
\end{abstract}
\section{Introduction}
Pedestrian emergency evacuation is a movement of people from a place of danger to a safer place in case of life-threatening incidents such as fire and terrorist attacks. Numerical simulation has been a popular approach to perform pedestrian emergency evacuation studies, for instance, predicting total evacuation time in a class room~\cite{Guo_TrB2012} and preparing an optimal evacuation plan for a large scale pedestrian facility~\cite{Abdelghany_EJOR2014}. 

Based on numerical simulations, it has been identified that evacuees are often in conflict with others when more than two evacuees try to move to the same position~\cite{Yanagisawa_PRE2007}. Game theory has been used to model strategic interactions among evacuees in such a conflict. Under game theoretic assumptions, each evacuee has his own strategies and selects a strategy in a way to maximize his own payoff. Various emergency evacuation simulations have been performed based on different game theory models including evolutionary game~\cite{Hao_PRE2011}, snowdrift game~\cite{Shi_PRE2013}, and spatial game~\cite{Heliovaara_PRE2013,vonSchantz_PRE2015}.

Although those game theory models successfully modeled evacuees’ egress, especially from a room, other aspects of evacuees’ behavior such as helping behavior have not been sufficiently studied. In the context of emergency evacuations, it has been reported that evacuees help injured evacuees to evacuate from the place of danger, for instance the WHO concert disaster occurred on December 3, 1979 in Cincinnati, Ohio, United States~\cite{Johnson_1987} and 2005 London bombings in United Kingdom~\cite{Drury_2009}. 

A few studies have investigated helping behavior in emergency evacuation by means of pedestrian simulation. Von Sivers~\textit{et al.}~\cite{vonSivers_PED2014,vonSivers_SafetySci2016} applied social identity and self-categorization theories to pedestrian simulation in order to simulate helping behavior observed in 2005 London bombings. In their studies, they assumed that all the evacuees share the same social identify which makes them be willing to help others rather than be selfish.  Lin and Wong~\cite{Lin_PRE2018} applied the volunteer’s dilemma game~\cite{Diekmann_JCR1985,Diekmann_SF2016} to model the behavior of volunteers who removed obstacles from the exit. Their work can be considered as a helping behavior modeling study in that some evacuees were voluntarily removing the obstacles so they helped others in the same room to evacuate faster.

One can observe that such a helping behavior provides a collective good in case of emergency evacuations. This is especially true when there are not enough rescuers, more injured persons can be rescued with the help of other evacuees than by only the rescuers. In order to study the collective effects of helping behavior in emergency evacuations, we have developed an agent-based model simulating such helping behaviors among evacuees. Based on the agent based model, we represent individual behaviors with a set of behavioral rules and then systematically study collective dynamics of interacting individuals. In our agent based model, we assumed that helping an injured person can be a costly behavior because the volunteer spends extra time and take a risk to assist the injured person in the evacuation. If individuals feel that helping behavior is a costly behavior for them, they might not turn into volunteers. Thus we implemented the volunteer’s dilemma game model~\cite{Diekmann_JCR1985,Diekmann_SF2016} to reflect the cost of helping behavior. Pedestrian movement is simulated based on social force model~\cite{Helbing_PRE1995}.

The remainder of this paper is organized as follows. The simulation model and its setup are explained in Sec.~\ref{section:method}. We then present its numerical simulation results with a phase diagram in Sec.~\ref{section:results}. Finally, we discuss the findings of this study in Sec.~\ref{section:conclusion}.

\section{Method}
\label{section:method}

\subsection{Volunteer's Dilemma Game}
\label{section:VDG}
We employ the volunteer’s dilemma game model to study helping behavior of passersby in a room evacuation~\cite{Diekmann_JCR1985,Diekmann_SF2016}. A passerby is an evacuee who is not injured and can play the volunteer’s dilemma game model. According to the volunteer’s dilemma game, two types of players are considered. Passerby $i$ can be either a volunteer (C) who helps an injured person to evacuate or a bystander (D) who does not help the injured person. Once a passerby decides to be a volunteer, he approaches to and then rescues the injured person. We can express the payoff of player $i$ in terms of collective good $U$ and volunteering cost $K < U$, see Table~\ref{table:payoff_matrix}. The payoff of a bystander (D) is $U$ if there is at least one volunteer, 0 otherwise. It can be understood that, bystanders are benefited by the volunteer. However, if nobody volunteers, the collective good $U$ cannot be produced because all the players are bystanders. The collective good $U$ can be produced by volunteers when they rescue injured persons. For simplicity, we assume that the value of $U$ is constant if there is at least one volunteer. The payoff of a volunteer (C) is always $U-K$, indicating that his payoff is constant regardless other players' choice. 

%
\begin{table}[]
\normalsize
\setlength{\tabcolsep}{6pt}
\centering
\caption{Payoff of a volunteer (C) and a bystander (D) for the different number of other players choosing C (based on Refs.~\cite{Diekmann_JCR1985,Diekmann_SF2016}). Here, $U$ is the collective good, $K < U$ is the volunteering cost, and $N \geq 2$ is the number of players.}
\label{table:payoff_matrix}
\resizebox{12cm}{!}{
\begin{tabular}{c*{6}{c}}
	\hline\hline 
	\multirow{2}{*}{Player $i$'s choice}& \multicolumn{5}{c}{The number of other players choosing C}\\
										& 0 & 1 & 2 & ... & $N-1$\\
	\hline
	Volunteer (C) & $U-K$	& $U-K$	& $U-K$	& $U-K$	& $U-K$	\\
	Bystander (D) & 0		& $U$	& $U$	& $U$	& $U$	\\
	\hline\hline
\end{tabular}
}
\end{table}

Actor $i$'s expected payoff $E_i$ is given as:
\begin{eqnarray}\label{eq:payoff_expected}
E_i = q_i \left( 1-\prod_{j \neq i}^{N} q_j \right)U + (1-q_i)(U-K).
\end{eqnarray}
\noindent
Here, $q_i$ is the probability that player $i$ chooses D and $1-q_i$ for choosing C. The number of players is indicated by $N$. The probability that all players $j \neq i$ choose D is denoted by $\prod q_i$ and $1-\prod q_i$ indicates the probability that at least one actor $j \neq i$ chooses C. The first term on the right hand side reflects the payoff of player $i$ when he selects D but benefited when there is at least one volunteer. The second term on the right hand side indicates the payoff of player $i$ if he selects C.

We assume that player $i$ adopts the mixed-strategy which is the best strategy for him. In a mixed-strategy equilibrium, every action played with positive probability must be a best response to other players’ mixed strategies. This implies that player $i$ is indifferent between choosing C and D, so a small change in the payoff $E_i$ with respect to $q_i$ (i.e., the probability of choosing D) becomes zero: 
\begin{eqnarray}\label{eq:mixed-strategy-equilibrium}
\frac{dE_i}{dq_i} = -U \prod_{j \neq i}^{N} q_j + K = 0.
\end{eqnarray}
\noindent
After assuming $q_i = q_j$, we can obtain probability that player $i$ chooses D 
\begin{eqnarray}\label{eq:q_i}
q_i = \left[ \frac{K}{U} \right]^{\frac{1}{N-1}} = \beta^{\frac{1}{N-1}},
\end{eqnarray}
\noindent
where $\beta = K/U$ is cost ratio, which can be interpreted as the risk of volunteering. Accordingly, the probability that player $i$ chooses C is given as 
\begin{eqnarray}\label{eq:p_i}
p_i = 1-q_i = 1-\beta^{\frac{1}{N-1}}.
\end{eqnarray}
\noindent
The probability that at least one player selects C is denoted by $p^{*}$, i.e., 
\begin{eqnarray}\label{eq:p_star}
p^{*}= 1-q_{i}^{N} = 1-\beta^{\frac{N}{N-1}}.
\end{eqnarray}

\begin{figure}[!t]
\centering
\begin{tabular}{cc}
	\includegraphics[width=.49\columnwidth]{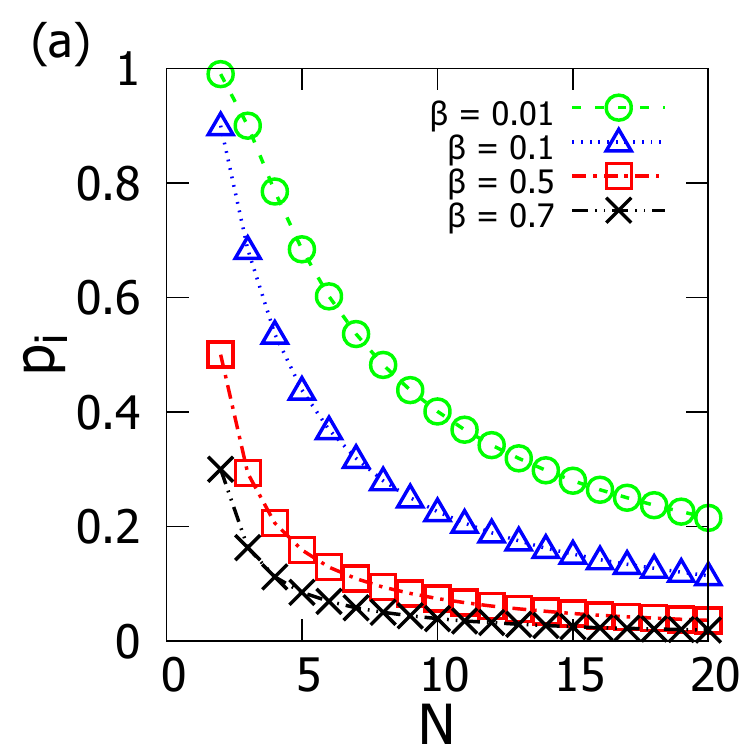}&
	\includegraphics[width=.49\columnwidth]{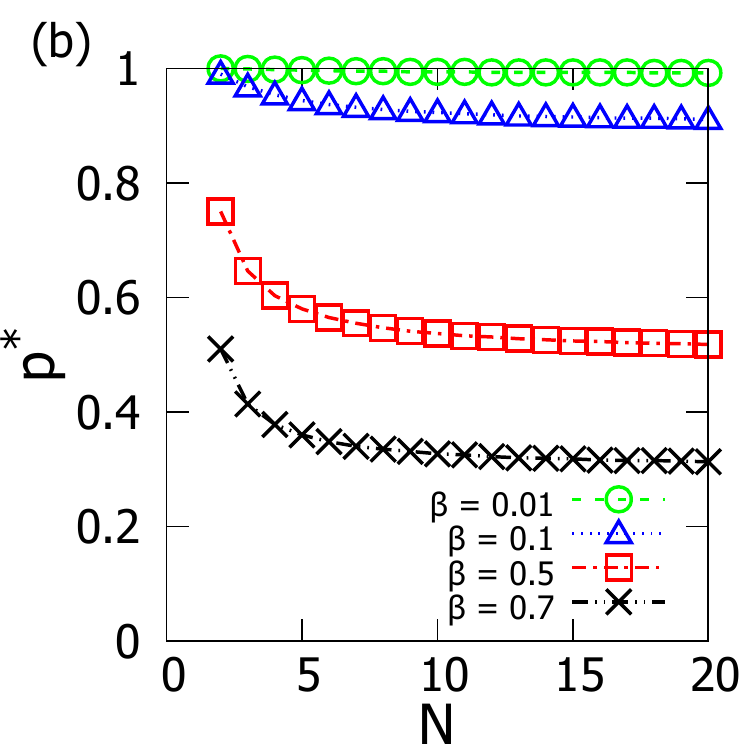}\\
\end{tabular}
\caption{Bystander effect on helping behavior: (a) $p_i$, the probability that player $i$ volunteers to rescue an injured person and (b) $p^{*}$, the probability that an injured person is rescued.}
\label{fig:probability} 
\end{figure}

Equations~\ref{eq:p_i} and~\ref{eq:p_star} show good agreement with the bystander effect, see Fig.~\ref{fig:probability}. Figure~\ref{fig:probability}(a) shows a decreasing trend of $p_i$ as the number of players $N$ increases, inferring that players are less likely to volunteer seemingly because they believe other players will volunteer. Note that the social pressure from other players is not considered here, so the existence of volunteers does not affect on players' behavior. Figure~\ref{fig:probability}(b) presents the trend of $p^{*}$ which reflects the chance that an injured person is rescued. As the number of players $N$ increases, the value of $p^{*}$ approaches to a certain value, $1-\beta$. 

\subsection{Social Force Model}
\label{section:SFM}
According to the social force model~\cite{Helbing_PRE1995}, the position and velocity of each pedestrian $i$ at time $t$, denoted by $\vec{x}_i(t)$ and $\vec{v}_i(t)$, evolve according to the following equations:
\begin{eqnarray}\label{eq:SFM_v}
\frac{\mathrm{d} \vec{x}_i(t)}{\mathrm{d} t} = \vec{v}_i(t)
\end{eqnarray}
\noindent 
and
\begin{eqnarray}\label{eq:SFM_acc}
\frac{\mathrm{d} \vec{v}_i(t)}{\mathrm{d} t} = \vec{f}_{i, d} + \sum_{j\neq i}^{ }{\vec{f}_{ij}} + \sum_{B}^{ }{\vec{f}_{iB}}.
\end{eqnarray}
\noindent
In Eq.~(\ref{eq:SFM_acc}), the driving force term $\vec{f}_{i, d} = (v_d\vec{e}_i - \vec{v}_i)/\tau$ describes the tendency of pedestrian $i$ moving toward his destination. Here, $v_d$ is the desired speed and $\vec{e}_i$ is a unit vector indicating the desired walking direction of pedestrian $i$. The relaxation time $\tau$ controls how quickly the pedestrian adapts one’s velocity to the desired velocity. The repulsive force terms $\vec{f}_{ij}$ and $\vec{f}_{iB}$ reflect his tendency to keep certain distance from other pedestrian $j$ and the boundary $B$, e.g., wall and obstacles. A more detailed description of the social force model can be found in previous studies~\cite{Helbing_PRE1995,Johansson_2007,Kwak_PRE2013,Viswanathan_EPJB2013}.

\subsection{Numerical Simulation Setup}
\label{section:setup}

\begin{figure}[!htp]
	\centering
	\includegraphics[width=9cm]{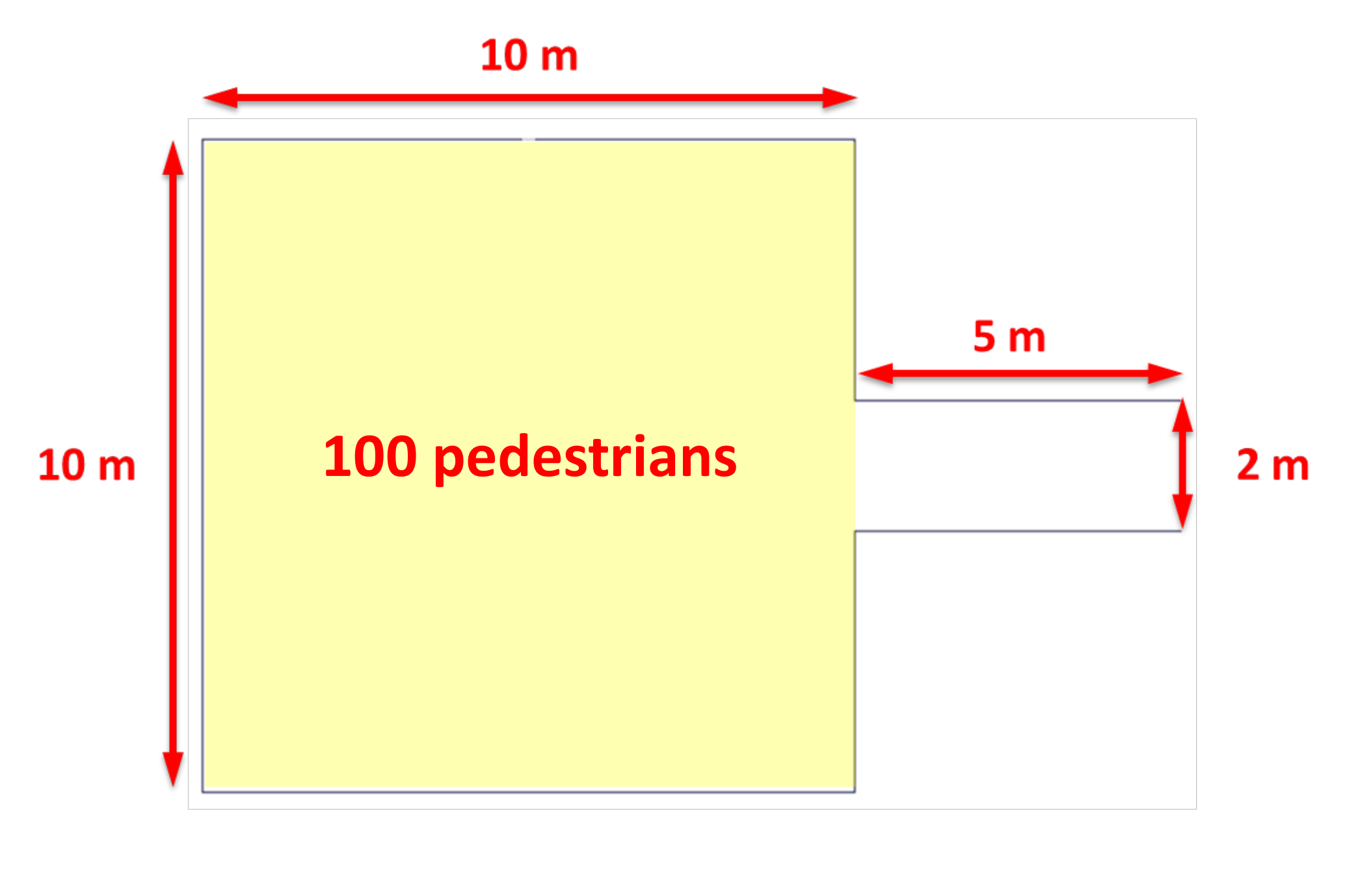}
	\caption{Schematic depiction of the numerical simulation setup. 100 pedestrians are placed in a 10m$\times$10m room indicated by a yellow shade area. Pedestrians are leaving the room through an exit corridor which is 5~m long and 2~m wide. The place of safety is set on the right, outside of the exit corridor.} 
	\label{fig:room_layout}
\end{figure}

Our agent-based model consists of helping behavior model and movement model. The helping behavior model computes the probability that a passerby would help an injured person based on the volunteer's dilemma game. The movement model calculates the sequence of pedestrian positions for each simulation time step. Our agent-based model was implemented from scratch in C++.

Each pedestrian is modeled by a circle with radius $r_i = 0.2$~m. $N_0 = 100$ pedestrians are placed in a 10m$\times$10m room indicated by a yellow shade area in Fig.~\ref{fig:room_layout}. Pedestrians are leaving the room through an exit corridor which is 5~m long and 2~m wide. The place of safety is set on the right, outside of the exit corridor. There are $N_i$ injured persons who need a help in escaping the room and $N = N_0 - N_i$ passersby who are ambulant. Some passersby might turn into volunteers who are going to approach to and then rescue the injured persons. The number of volunteers is determined based on the volunteer's dilemma game presented in Sec.~\ref{section:VDG}.  

The volunteer's dilemma game is updated for each second. We assumed that the volunteer's dilemma game is a macroscopic behavior like goal selection and path navigation patterns~\cite{Zhong_AAMAS2016}. In line with Heli\"{o}vaara \textit{et al.}~\cite{Heliovaara_PRE2013}, each passerby can play the volunteer's dilemma game a few times during the whole simulation period. With the update frequency of one time per second, most of passersby play the volunteer's dilemma game up to ten times before they leave the room. A passerby can decide whether he will volunteer to rescue an injured person within a range of 3~m. Once the volunteer decides to rescue the injured person, then he shifts his desired direction walking vector $\vec e_i$ toward the position of injured person. Once the volunteer reaches the injured person, he will flee to the place of safety with the injured person after a preparation time of 5~s. 

The pedestrian movement is updated with the social force model in Eq.~(\ref{eq:SFM_acc}). The passersby move with the initial desired speed $v_d = v_{d, 0}=1.2$~m/s and with relaxation time $\tau = 0.5$~s, and their speed cannot exceed $v_{\rm max} = 2.0$~m/s. Until now, the speed of volunteers rescuing the injured persons is often assumed by the modelers, like the work of Von Sivers~\textit{et al.}~\cite{vonSivers_PED2014,vonSivers_SafetySci2016}. We applied speed reduction factor $\alpha = 0.5$ to the volunteers rescuing the injured persons, so they move with a reduced desired speed $v_{d} = \alpha v_{d, 0} = 0.6$~m/s. Following previous studies~\cite{Zanlungo_EPL2011,Zanlungo_PRE2014,Kwak_PRE2017}, we discretized the numerical integration of Eq.~(\ref{eq:SFM_acc}) using the first-order Euler method:
\begin{eqnarray}\label{eq:Euler_method}
\vec{v}_i(t + \Delta t) &=& \vec{v}_i(t) + \vec{a}_i(t)\Delta t,\\
\vec{x}_i(t + \Delta t) &=& \vec{x}_i(t) + \vec{v}_i(t + \Delta t)\Delta t.
\end{eqnarray}
\noindent
Here, $\vec{a}_i(t)$ is the acceleration of pedestrian $i$ at time $t$ which can be obtained from Eq.~(\ref{eq:SFM_acc}). The velocity and position of pedestrian $i$ is denoted by $\vec{v}_i(t)$ and $\vec{x}_i(t)$, respectively. The time step $\Delta t$ is set as 0.05~s.

\section{Results and Discussion}
\label{section:results}
%
%

\begin{figure}[!htp]
	\centering
	\includegraphics[width=\textwidth]{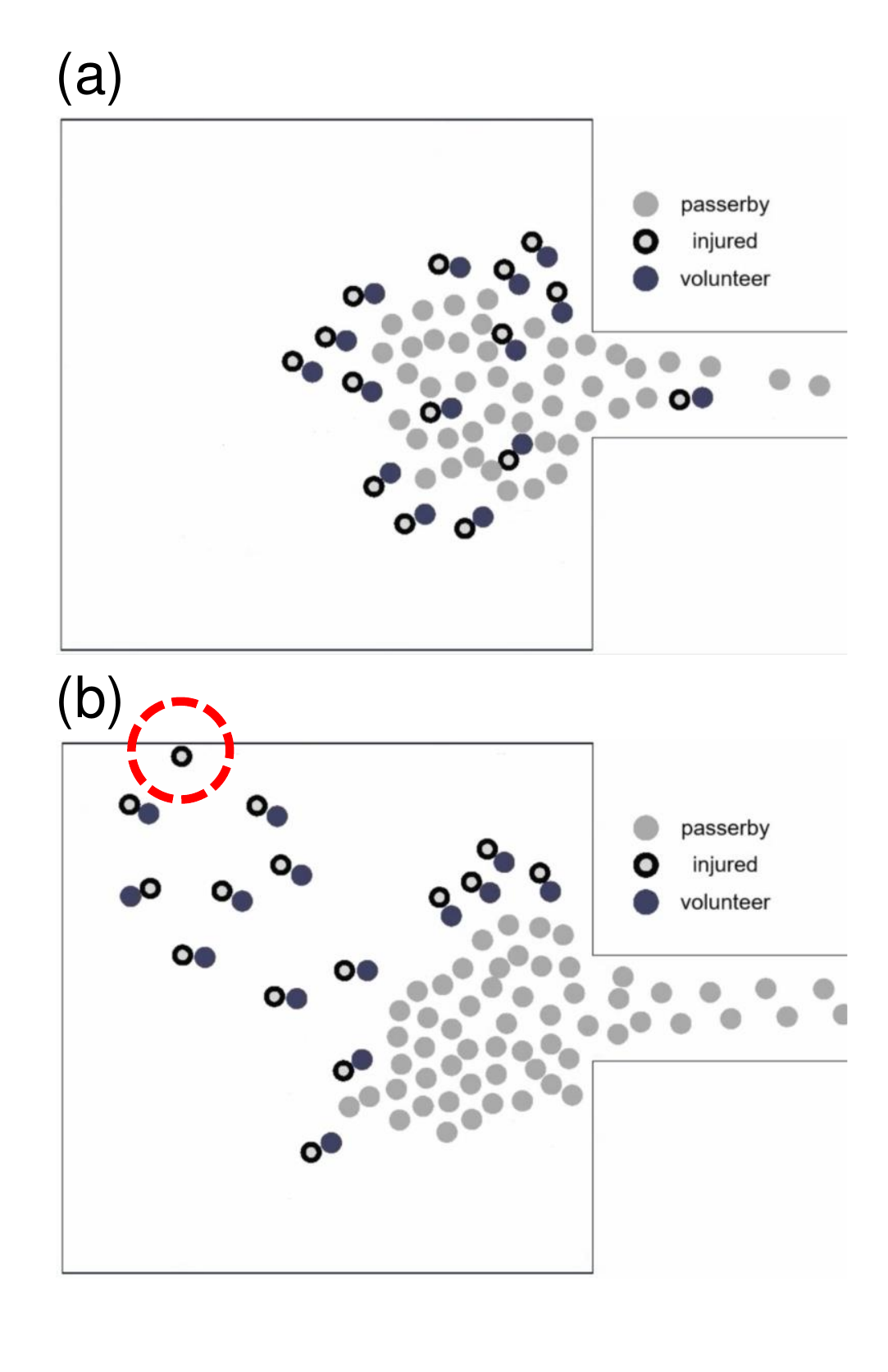}
	\caption{Snapshots of helping behavior in a room evacuation scenario: (a) all the injured persons are rescued in case of $N_i = 15$ and $\beta = 0.1$, and (b) some injured persons are not rescued (in the red dotted circle) in case of $N_i = 15$ and $\beta = 0.2$. Open black circles indicate injured persons and full dark circles show volunteers helping the injured persons. Gray circles represent the passersby.}
	\label{fig:snapshots}
\end{figure}

\begin{figure}[!htp]
	\centering
	\includegraphics[width=8cm]{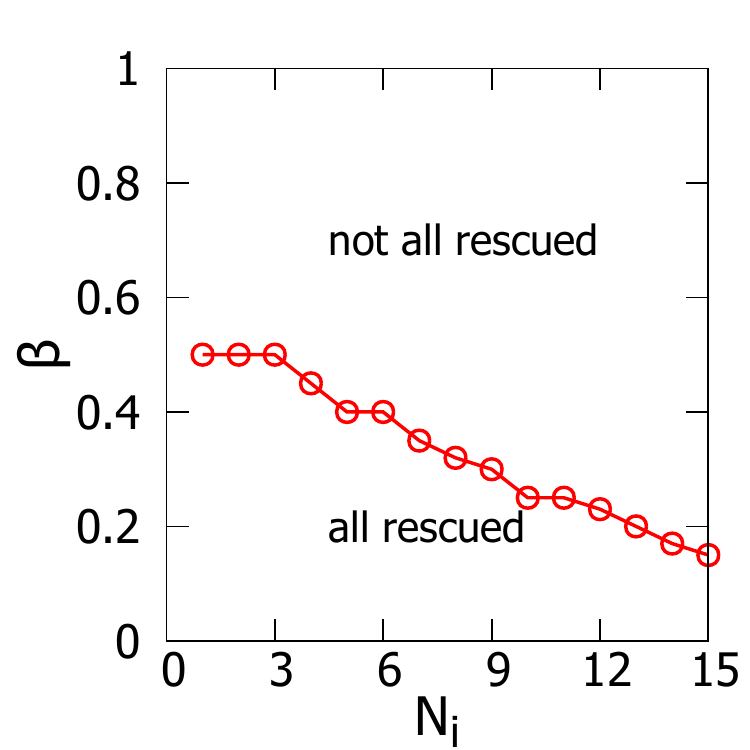}
	\caption{Schematic phase diagram of collective helping behavior in the room evacuation scenario.} \label{fig:phase_diagram}
\end{figure}

Figure~\ref{fig:snapshots} shows snapshots of our agent-based model simulations. Open black circles indicate injured persons and full dark circles show volunteers helping the injured persons. Gray circles represent the passersby. If the helping cost is low, all the injured persons are likely to be rescued, as shown in Fig.~\ref{fig:snapshots}(a). However, if the helping cost is high (i.e., high $\beta$), then some injured persons might not be rescued, see the red dotted circle in Fig.~\ref{fig:snapshots}(b). By systematically changing the value of $N_i$ and $\beta$, we observed different patterns of collective helping behaviors summarized in the schematic phase diagram (see Fig.~\ref{fig:phase_diagram}). For each parameter combination ($N_i$, $\beta$), we performed 30 independent simulation runs.

\begin{figure}[!htp]
	\centering
	\includegraphics[width=8cm]{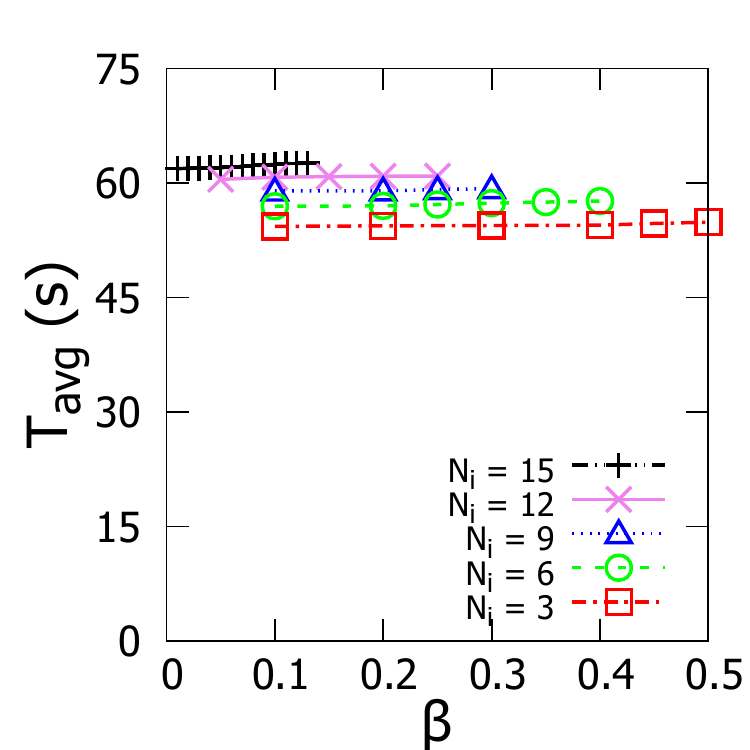}
	\caption{Average total evacuation time $T_{\rm avg}$ as a function of the number of injured persons $N_i$ and $\beta$.} 
	\label{fig:total_evac_time_beta}
\end{figure}

\begin{figure}[!t]
	\centering
	\begin{tabular}{cc}
		\includegraphics[width=.49\columnwidth]{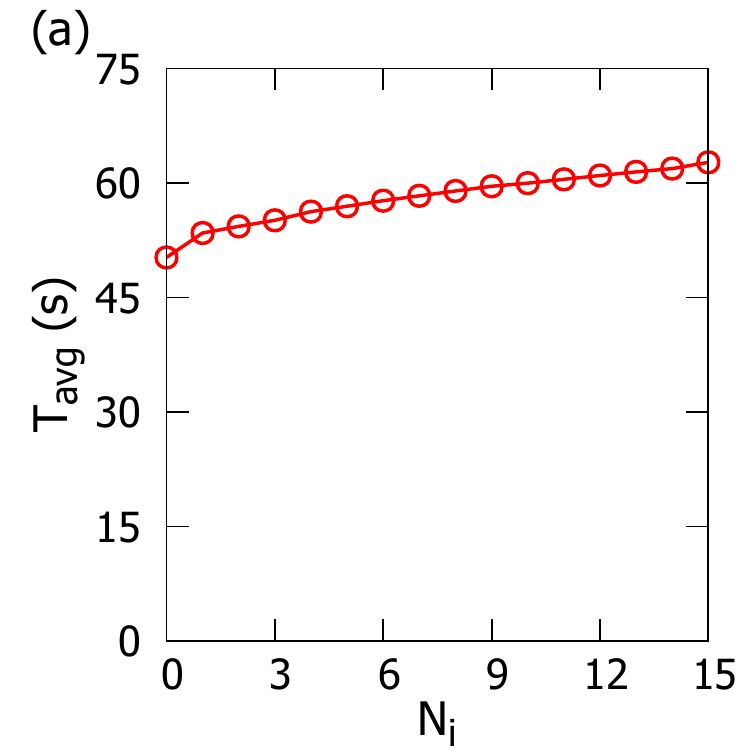}&
		\includegraphics[width=.49\columnwidth]{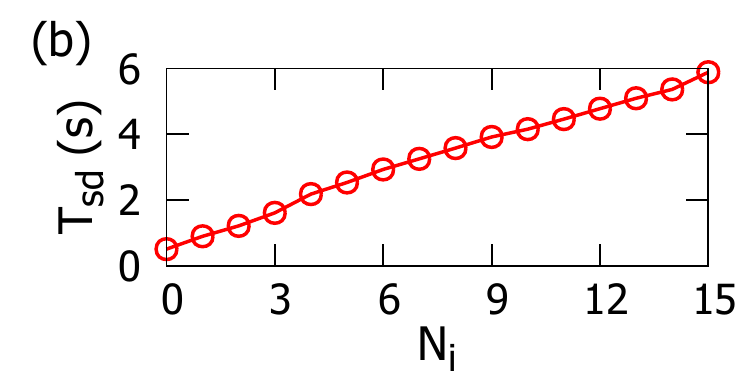}\\
	\end{tabular}
	\caption{Total evacuation time in case of $\beta = 0.1$ against the number of injured persons $N_i$: (a) average $T_{\rm avg}$ and (b) standard deviation $T_{\rm sd}$.}
	\label{fig:total_evac_time_injured}
\end{figure}

We also looked into the impact of cost ratio $\beta$ and the number of injured persons $N_i$ on the total evacuation time $T$. The total evacuation time $T$ is defined as the length of period from the start of evacuation to the moment when the last evacuee leaves the exit corridor. We measured the average and standard deviation of the total evacuation time, i.e., $T_{\rm avg}$ and $T_{\rm sd}$, based on the values of total evacuation time $T$ obtained over 30 independent simulation runs for each parameter combination ($N_i$, $\beta$). Figure~\ref{fig:total_evac_time_beta} indicates that change in the value of $\beta$ does not make a noticeable different to the average total evacuation time $T_{\rm avg}$. This is seemingly because $\beta$ only affects the probability that a passerby turns into a volunteer. As indicated in Fig.~\ref{fig:total_evac_time_injured}(a), the average total evacuation time $T_{\rm avg}$ increases as the number of injured persons $N_i$ grows. Having more injured persons indicates that there are more volunteers who move in the reduced desired speed, so the total evacuation time increases due to the volunteers rescuing the injured persons. In addition, the standard deviation of total evacuation time $T_{\rm sd}$ increases as the number of injured persons $N_i$ grows, in that the difference in evacuation time among evacuees gets larger.

\section{Conclusion}
\label{section:conclusion}
We have numerically investigated helping behavior among evacuees in a room evacuation scenario. Our simulation model is based on the volunteer's dilemma game reflecting volunteering cost and social force model simulating pedestrian movement. We characterized collective helping behavior patterns by systematically controlling the values of cost ratio $\beta$ and the number of injured pedestrians $N_i$. For low cost ratio values, one can expect that all the injured pedestrians are rescued by volunteers. For high cost ratio values, on the other hand, it was observed that not all the injured persons can be rescued. When the number of injured persons is large, a low value of cost ratio yields a result that all the injured pedestrians are rescued. A schematic phase diagram summarizing the collective helping behavior patterns is presented.

A very simple room evacuation scenario has been used in order to study the fundamental role of helping behavior in the evacuation especially the number of evacuated pedestrians. In this study, the severity of injuries are assumed to be the same for all the injured persons, so each injured person can be rescued by a volunteer. According to patient triage scale in Singapore~\cite{Parker_2019,SGH_AcuityScale}, patients can be categorized based on the severity of injuries and the desired number of volunteers are different for various types of injuries. Future work can reflect the impact of patient injury levels in collective helping behavior by assuming different number of required volunteers for each patient. This study can be extended from the perspective of game theory. As stated in Diekmann's study~\cite{Diekmann_SF2016}, it can be interesting to introduce different values of collective good $U$ and volunteering cost $K$ to each passerby. By doing that, we can reflect personal difference in willingness to volunteer in emergency evacuations. In addition, one can imagine that the value of U can be changed depending on the number of injured persons and volunteers. For instance, the values of $U$ are different when there are one injured person, one volunteer and two injured persons, three volunteers. Evolutionary game~\cite{Hao_PRE2011} can be also introduced in order to reflect behavioral changes of passersby influenced by the existence of volunteers, which might be observable in emergency evacuations. 

\section*{Acknowledgements}
This research is supported by National Research Foundation (NRF) Singapore, GOVTECH under its Virtual Singapore program Grant No. NRF2017VSG-AT3DCM001-031.  

%
%
%
%

\end{document}